\documentstyle[aps,prl]{revtex}

\begin{document}
\title{ Nonextensive statistical mechanics: A brief review of its present status}
\author{C. Tsallis \thanks{tsallis@cbpf.br}  }
\address{Centro Brasileiro de Pesquisas F\'{\i}sicas, 
Xavier Sigaud 150, 22290-180, Rio de Janeiro, Brazil\\
Centro de Fisica da Materia Condensada, Universidade de Lisboa, Av. Prof. Gama Pinto 2, P-1649-003 Lisboa, Portugal}
\draft
\maketitle
\begin{abstract}
We briefly review the present status of nonextensive statistical mechanics. We focus on (i) the central equations of the formalism, (ii) the most recent applications in physics and other sciences,  (iii) the {\it a priori} determination (from microscopic dynamics) of the entropic index $q$  for two important classes of physical systems, namely
low-dimensional maps (both dissipative and conservative) and long-range interacting many-body hamiltonian classical systems. 
\end{abstract}

\date{\today}

\section{Central equations of nonextensive statistical mechanics}

Nonextensive statistical mechanics and thermodynamics were introduced in 1988 \cite{tsallis}, and further developed in 1991 \cite{curadotsallis} and 1998 \cite{tsallismendesplastino}, with the aim of extending the domain of applicability of statistical mechanical procedures to systems where Boltzmann-Gibbs (BG) thermal statistics and standard thermodynamics present serious mathematical difficulties or just plainly fail. Indeed, a rapidly increasing number of systems are pointed out in the literature for which the usual functions appearing in BG statistics appear to be violated. Some of these cases are satisfactorily handled within the formalism we are here addressing (see 
\cite{review} for reviews and \cite{biblio} for a regularly updated bibliography which includes crucial contributions and clarifications that many scientists have given along the years). Let us start by just reminding the central equations.

First of all, the exponential function $e^x$ is generalized into the {\it $q$-exponential function}
\begin{equation}
e_q^x \equiv [1+(1-q)\;x]^{\frac{1}{1-q}} \;\;\;\;(q \in {\cal R}) \;.
\end{equation}
We can trivially verify that this (nonnegative and monotonically increasing) function (i) for $q \to 1$ yields $e _1^x=e^x$ ($\forall x$), (ii) for $q>1$, vanishes as a power-law when $x \to  -\infty$ and diverges at $x=1/(q-1)$, and (iii) for $q<1$, has a cutoff at $x=-1/(1-q)$, below which it is defined to be identically zero. If $x \to 0$ we have $e_q^x \sim 1+x \; (\forall q)$.

The inverse function of the $q$-exponential is the {\it $q$-logarithm}, defined as follows:
\begin{equation}
\ln_q  x \equiv \frac{x^{1-q} -1}{1-q} \;\;\;\;(q \in {\cal R}) \;.
\end{equation}
Of course $\ln_1x = \ln x \; (\forall x)$. If $x \to 1$ we have $\ln_q x \sim x-1 \; (\forall q)$.

The nonextensive entropic form we postulate is
\begin{equation}
S_q =k \frac{1- \sum_{i=1}^W p_i^q}{q-1} \;\; \;\;\; (\sum_{i=1}^W p_i=1;  \; q \in {\cal R}) \;,
\end{equation}
where $W$ is the total number of microscopic configurations, whose probabilities are $\{p_i\}$. Without loss of generality we shall from now on assume $k=1$. We can verify that, for $q \to 1$, this entropy reproduces the usual Boltzmann-Gibbs-Shannon one, namely $S_1=-\sum_{i=1}^Wp_i \ln p_i$. The continuous and the quantum expressions of $S_q$ are respectively given by
\begin{equation}
S_q = \frac{1- \int dx\;[p(x)]^q}{q-1} 
\end{equation}
and
\begin{equation}
S_q = \frac{1- Tr \rho^q}{q-1} \;,
\end{equation}
where $\rho$ is the matrix density. Unless specifically declared in what follows, we shall be using the form of Eq. (3). It is easy to verify that all its generic properties can be straightforwardly adapted to both the continuous and quantum cases.

$S_q$ can be written as
\begin{equation}
S_q = \langle  \ln_q \frac{1}{p_i} \rangle,
\end{equation}
where the {\it expectation value} $\langle (...) \rangle \equiv \sum_{i=1}^W p_i(...)$. It can also be written as
\begin{equation}
S_q =\langle - \ln_q p_i \rangle_q \;,
\end{equation}
where the {\it unnormalized $q$-expectation value} is defined to be $\langle (...) \rangle_q \equiv \sum_{i=1}^W p_i^q(...)$. Of course 
$\langle (...) \rangle_1 =\langle (...) \rangle$. This is a good point for defining also the {\it normalized $q$-expectation value} 
\begin{equation}
\langle \langle (...) \rangle \rangle_q \equiv     \frac{\sum_{i=1}^W p_i^q(...)  }{\sum_{i=1}^Wp_i^q } \;,
\end{equation}
which naturally emerges in the formalism. We verify trivially that $   \langle \langle (...) \rangle \rangle_1 = \langle (...) \rangle_1= \langle (...) \rangle $, and also that $   \langle \langle (...) \rangle \rangle_q = \langle (...) \rangle_q / \langle 1 \rangle_q $.

If $A$ and $B$ are two independent systems (i.e., $p_{ij}^{A+B}=p_i^A p_j^B \;\forall(i,j)$), then we have that
\begin{equation}
S_q(A+B)=S_q(A)+S_q(B)+(1-q)S_q(A)S_q(B).
\end{equation}
It is from this property that the name {\it nonextensive} statistical mechanics was coined. The cases $q<1$ and $q>1$ respectively correspond to {\it superextensivity} and {\it subextensivity} of $S_q$ since in all cases $S_q \ge 0$. 

At equiprobability, i.e., $p_i=1/W$, we obtain
\begin{equation}
S_q=\ln_q W\;,
\end{equation}
which is the basis for the microcanonical ensemble.

For the {\it thermal equilibrium} corresponding to the canonical ensemble of a Hamiltonian system, we optimize $S_q$ with the constraints $\sum_{i=1}^Wp_i=1$ and $\langle \langle \epsilon_i \rangle \rangle_q=U_q$, where $\{\epsilon_i\}$ are the eigenvalues of the Hamiltonian of the system, and $U_q$ is the generalized internal energy. We obtain \cite{tsallismendesplastino}
\begin{equation}
p_i=\frac{e_q^{-\beta(\epsilon_i-U_q)}}{Z_q}  \propto \frac{1}{[1+(q-1) \beta (\epsilon_i-U_q)]^{\frac{1}{q-1}}} \propto   \frac{1}{[1+(q-1) \beta^{\prime} \epsilon_i]^{\frac{1}{q-1}}}    \;,
\end{equation}
where $\beta$ is the Lagrange parameter, $Z_q \equiv \sum_{j=1}^W e_q^{-\beta (\epsilon_j-U_q)}$ and $\beta^\prime$ a well defined function of $\beta$. For $q=1$ we recover the celebrated BG weight. When $\beta>0$ and the energy $\epsilon_i$ increases, the probability decays like a power law for $q>1$ and exhibits a cutoff for $q<1$.

Analogously, if we optimize $S_q$ as given by Eq. (4) with the constraints $\int dx\;p(x)=1$ and $\langle \langle x^2 \rangle \rangle_q=\sigma^2$ ($\sigma>0$), we obtain  the $q$-generalization of the Gaussian distribution, namely \cite{levy}
\begin{equation}
p_q(x)=\frac{e_q^{-{\bar \beta} x^2}}{ \int dy\;  e_q^{-{\bar \beta} y^2}   }  \propto \frac{1}{[1+(q-1) {\bar \beta} x^2]^{\frac{1}{q-1}} }      \;\;\;\;(q<3), 
\end{equation}
where ${\bar \beta}$ can be straightforward and explicitly related to $\sigma$. The variance of these distributions is finite if $q<5/3$ and diverges if $5/3<q<3$. For $q=2$ we have the Lorentzian distribution. For $q \ge 3$ the function is not normalizable, and therefore is unacceptable as a distribution of probabilities.
 
Let us now address typical time dependences. Let us assume that $\xi(t)$ is a quantity characterizing an exponential behavior and satisfying $\xi(0)=1$. Such is the typical case for the sensitivity to the initial conditions of a one-dimensional chaotic system, where $\xi(t) \equiv \lim_{\Delta x(0) \to 0} [\Delta x(t)/\Delta x(0)]$, where $\Delta x(t)$ is the discrepancy at time $t$ of two trajectories which, at $t=0$, started at $x(0)$ and at $x^{\prime}(0)$. Another example is a population which relaxes to zero. If $N(t)$ is the number of elements, then we can define $\xi(t) \equiv N(t)/N(0)$. The quantity $\xi(t)$ monotonically increases in our first example (sensitivity), whereas it decreases in the second one (relaxation).
The basic equation that $\xi$ satisfies is generically 
\begin{equation}
{\dot \xi} = \lambda_1 \;\xi\;,
\end{equation}
hence $\xi(t) = e^{\lambda_1t}$. In our example of the chaotic system, $\lambda_1$ is the Lyapunov exponent. In our population example, we have $\lambda_1 \equiv -1/\tau_1$, where $\tau_1$ is the relaxation time. What happens in the marginal case $\lambda_1=0$? Typically we have
\begin{equation}
{\dot \xi} = \lambda_q \;\xi^q\;,
\end{equation}
hence
\begin{equation}
 \xi = e_q^{\lambda_qt} =[1+(1-q) \lambda_q\;t]^{\frac{1}{1-q}} = \frac{1}{    [1+(q-1)(- \lambda_q)\;t]^{\frac{1}{q-1}}  } \;.
\end{equation}
This quantity monotonically increases if $\lambda_q>0$ and $q<1$, and decreases if $\lambda_q \equiv -1/\tau_q <0$ and $q>1$. In both cases it does so as a power law, instead of exponentially. In the limit $t \to 0$, we have $\xi \sim 1+\lambda_q\;t \; (\forall q)$.

A more general situation might occur when both $\lambda_1$ and $\lambda_q$ are different from zero. In such case, many phenomena will  be described by the following differential equation:
\begin{equation}
{\dot \xi} = \lambda_1 \; \xi + \lambda_q \;\xi^q\;,
\end{equation}
hence
\begin{equation}
\xi = \Bigl[1- \frac{\lambda_{q}}{\lambda_1} + \frac{\lambda_{q}}{\lambda_1}e^{(1-q)\lambda_1 t} \Bigr]^{\frac{1}{1-q}} \;.
\end{equation}
If $q<1$ and $0<\lambda_1 << \lambda_q$, $\xi$ increases linearly with $t$ for small times, as $t^{\frac{1}{1-q}}$ for intermediate times, and like $e^{\lambda_1t}$ for large times. If $q>1$ and $0>\lambda_1 >> \lambda_q$, $\xi$ decreases linearly with $t$ for small times, as $1/t^{\frac{1}{q-1}}$ for intermediate times, and like $e^{-|\lambda_1|t}$ for large times. 

\section{Applications in and out from equilibrium} 

A considerable amount of applications and connections have been advanced in the literature using, in a variety of manners, the above formalism. They concern physics, astrophysics, geophysics, chemistry, biology, mathematics, economics, linguistics, engineering, medicine, physiology, cognitive psychology, sports and others \cite{biblio}. It seems appropriate to say that the fact that the range of applications is so wide probably is deeply related to and reflects the ubiquity of self-organized criticality \cite{perbak}, fractal structures \cite{mandelbrot} and, ultimately, power laws in nature. In particular, a natural arena for this statistical mechanics appears to be the so called {\it complex systems} \cite{gellmann}.

We shall briefly review  here four recent applications, namely fully developed turbulence \cite{ramos,beck,arimitsu}, hadronic jets produced by electron-positron annihilation \cite{epjet}, motion of {\it Hydra viridissima} \cite{hydra}, and quantitative linguistics \cite{montemurro}. \\

\noindent
{\it Fully developed turbulence:} 

As early as in 1996 Boghosian made the first application of the present formalism to turbulence \cite{boghosian}. That was for plasma. What we shall instead focus on here is fully developed turbulence in normal fluids. Ramos et al advanced in 1999 \cite{ramos} the possibility of nonextensive statistical mechanics being useful for such systems. The idea was since then further developed by Beck \cite{beck} and by the Arimitsu's \cite{arimitsu}, basically simultaneous and independently. They proposed theories within which the probability distribution of the velocity differences and related quantities are deduced from basic considerations. We present in Fig. 1 the comparison of Beck's theoretical results with recent high precision experimental data for Lagrangian and Eulerian turbulences \cite{beck2}. In Fig. 2 we show an analogous comparison between Arimitsu's theoretical results and recent computer experimental data.  \\

\noindent
{\it Hadronic jets:}  

High energy frontal collisions of electron and positron annihilate both and produce relativistic hadronic jets. The distribution of the tranverse momenta of these jets admits, as advanced by Fermi, Feynman, Hagedorn and others, a thermostatistical theoretical approach. Hagedorn's 1965 theory was $q$-extended by Bediaga et al in 1999. Their results \cite{epjet}, as well as related ones by Beck \cite{epjet}, compare quite well with the available CERN experimental data, as can be seen in Fig. 3. As important as this is the fact that both Bediaga et al and Beck theories recover a crucial feature of Hagedorn's scenario, namely that the temperature to be associated with the distributions should {\it not} depend on the collision energy. \\ 

\noindent
{\it Hydra viridissima:}  

The motion of cell aggregates of {\it Hydra viridissima} in physiological solution was studied by Upadhyaya et al \cite{hydra}. The strongly nonmaxwellian distribution of velocities is quite well fitted with $q \simeq 1.5$: See Fig. 4. They also carried diffusion measurements and verified that diffusion is anomalous. Under the assumption that it is of the correlated type addressed in \cite{bukman}, they obtained once again $q \simeq 1.5$: See Fig. 5. In other words, two different experiments of motion were fitted by one and the same value for $q$.  \\

\noindent
{\it Linguistics:}  

The frequency $p$ of words used in a text (say a book or a set of books of one or more authors) as a function of their rank $r$ roughly follows the so called {\it Zipf law}, namely $p \propto 1/r$. This law was improved by Mandelbrot in the form $p \propto 1/(a+r)^\gamma$. This form fits better real data and is sometimes called the {\it Zipf-Mandelbrot law}. This law precisely is Eq. (15) with the notation $(t,\xi) \equiv (r,p)$ (or equivalently Eq. (11) where the energy plays the role of the word rank), as first argued by  Denisov in 1997 \cite{denisov}. However, although quite faithful at low and intermediate word ranks, the Zipf-Mandelbrot law fails at high ranks. This point has been addressed recently by Montemurro \cite{montemurro}: See Fig. 6, where results are shown from well known texts in English, Greek, Italian and Spanish, from authors of very different historical periods and literary styles (all reasonably well fitted by using $q \simeq 1.9$ hence $\gamma = 1/(q-1) \simeq 1.1$). Montemurro has shown that the fittings are all sensibly improved by using the present Eq. (17) instead of Eq. (15). The agreement obtained is illustrated with a large set of books by Dickens as shown in Fig. 7. In other words, at large values of the word rank a crossover occurs from $q \simeq 1.9$ to $q \simeq 1$. The reason for this interesting phenomenon is unknown. As a plausible hypothesis, we would like to advance that it might be related to the fact that most authors possibly use the very rare words in a manner which reflects their relatively poor knowledge of their exact meaning. This attitude could make those words to be used slightly uncorrelated with the context within which they are placed. It is however clear that this phenomenon is a very subtle one, and its full elucidation would presumably require very sophisticated analysis.

\section{A priori determination of the entropic index $q$}

Physical bounds to the admissible values of $q$ were first discussed as early as 1993 by Plastino and Plastino \cite{plastinospolytrope} (for self-gravitating systems), and since then by many others. However, the strict aim of the present Section is how is to be determined the value of $q$ to be associated with a specific system whose microscopic (or mesoscopic) dynamics is exactly known. This is to say, how the knowledge of the rules that provide the (continuous or discrete) time evolution of the system can be used in order to  determine without ambiguity the appropriate value(s) of $q$ for that system. This crucial question must be answered for the present proposal to be a complete theory, in the sense that it is in principle able to predict the results to be expected in all types of experiments with well defined systems. 
This question has by now been answered in several important classes of systems. We shall briefly review here two of them, namely low-dimensional maps and long-range many-body classical Hamiltonian systems. \\

\subsection{Low-dimensional maps}

We shall focus on one- and two-dimensional maps. The one-dimensional maps necessarily are dissipative. The two-dimensional ones can be either dissipative or conservative, but we shall primarily address the latter. Indeed, on one hand the dissipative two-dimensional maps are dynamically not so different from the one-dimensional ones. On the other hand, two-dimensional conservative maps provide us an {\it avant premiere} of Hamiltonian systems. \\

\noindent
{\it One-dimensional maps:}  

We consider here one-dimensional dissipative maps of the type 
\begin{equation}
x_{t+1} =f(x_t;a,z) \;\;\; (t=0,1,2,...; \; x_t \in [x_{min},x_{max}]) \;,
\end{equation}
where $a \in {\cal R}$ is a control parameter such that when it increases for fixed $z$, it makes the map to become chaotic (we note $a_c$ the smallest value of $a$ above which the system can be chaotic; $z \in {\cal R}$ is another control parameter which differs essentially from $a$ in the sense that $z$ controls the universality class of the chaotic attractor emerging at $a=a_c(z)$; the function $f$ is such that chaotic and nonchaotic behaviors are possible for the variable $x$, depending on the values of $(a,z)$. A paradigmatic such map is the so called {\it $z$-logistic map}, defined as
\begin{equation}
x_{t+1} =1-a|x_t|^z\;\;\; (t=0,1,2,...; \; x_t \in [-1,1]) \;,
\end{equation}
with $a \in (0,2]$ and $z>1$. The critical value  $a_c(z)$ ({\it chaos threshold} or {\it edge of chaos}) monotonically increases from 1 to 2 when $z$ increases from 1 to infinity; $a_c(2)=1.401155...$ . For $z=2$ this map is, as well known, isomorphic to $X_{t+1} \propto X_t(1-X_t)$.
The $z$-logistic maps and several others have already been studied \cite{maps1} within the nonextensive scenario. We briefly review here their main properties. Most of these properties have been found heuristically, and no theorems or rigorous results are available. Consequently, we are unable to precisely specify how generic are the properties we are going to describe. We know, however, that wide classes of maps do satisfy them.

Let us first address the sensitivity to the initial conditions. For all values of $a$ for which the Lyapunov exponent $\lambda_1$ is nonzero we verify  that $q=1$, i.e., $\xi(t) =e^{\lambda_1\;t}$, with $\lambda_1 <0$ for most values of $a < a_c$, and $\lambda_1 >0$ for most values of $a > a_c$. However, for the infinite number of values of $a$ for which $\lambda_1=0$ we verify that $q \ne1$. More precisely, for values of $a$ such as those for which bifurcations occur between finite cycle attractors of say the $z$-logistic map, we verify the validity of Eq. (15) with $q>1$ and $\lambda_q <0$ (this has been very recently proved \cite{fulvioalberto1}). 
For $a=a_c(z)$ we verify that $\xi(t)$ exhibits a complex behavior which has, nevertheless, a simple upper bound which satisfies Eq. (15) with $q<1$ (from now on noted $q_{sen}(z)$, where the subindex {\it sen} stands for {\it sensitivity}) and $\lambda_q(z) >0$ (also this has been very recently proved \cite{fulvioalberto2}) . For the universality class of the $z$-logistic map we verify that $q_{sen}$ monotonically increases from minus infinity to a value slightly below unity, when $z$ increases from 1 to infinity ($q_{sen}(2) = 0.2445...$). For the $z$-cercle and other maps we verify similar behaviors.

Let us now address the attractor in $x$ space for $a=a_c(z)$. Its anomalous geometry can be usefully characterized by the so called  {\it multifractal function} $f(\alpha,z)$ which typically is defined in the interval $ \alpha_{min}(z) \le \alpha \le \alpha_{max}(z)$, and whose maximal value is the {\it fractal} or {\it Hausdorff} dimension $d_f(z)$. For the $z$-logistic map universality class we have $d_f(z) <1$, whereas for the $z$-circle map universality class we have $d_f(z)=1 \;(\forall z)$. In all the cases we have checked, we verify a remarkable scaling law, namely \cite{lyratsallis}
\begin{equation}
\frac{1}{1-q_{sen}(z)} = \frac{1}{\alpha_{min}(z)} - \frac{1}{\alpha_{max}(z)} \;\;\; (\forall z).
\end{equation}
This relation has purely geometrical quantities at its right hand member, and a dynamical quantity at its left hand member. It can also be shown that
\begin{equation}
\frac{1}{1-q_{sen}(z)} = (z-1) \frac{\ln \alpha_F}{\ln b} \;\;\; (\forall z),
\end{equation}
where $\alpha_F$ is one of the two well known Feigenbaum constants, and $b$ is the attractor scaling ($b=2$ for period-doubling bifurcations; $b=2/(\sqrt{5}-1)$ for cercle maps).

Let us now address the entropy production $dS_q(t)/dt$. We first make a partition of the interval $[x_{min},x_{max}]$ into $W$ nonoverlaping little windows. We place (randomly or not) inside one of those $W$ windows a large number $N$ of initial conditions, and run the map $t$ times for each of these points. We generically verify that the $N$ points spread into the windows, in such a way that we have the set $\{N_i\}(t)$  (with $\sum_{i=1}^W N_i(t) = N \;, \forall t$ ) . With these numbers we can define the set of probabilities $\{p_i(t)\}$ where $p_i(t) \equiv N_i(t)/N \;( \forall i)$. We then choose a value for $q$ and calculate $S_q(t)$ by using Eq. (3). We then make an average $\langle S_q \rangle(t)$ over a few or many initial windows (see \cite{maps1,latorabarangerrapisardatsallis} for details), and finally evaluate numerically $\lim_{t \to \infty} \lim_{W \to \infty} \lim_{N \to \infty} \langle S_q\rangle (t)/t$. 
We verify a very interesting result \cite{latorabarangerrapisardatsallis}, namely that this limit is {\it finite} only for $q=q_{sen}(z)$; it diverges for all $q<q_{sen}(z)$ and vanishes for all $q>q_{sen}(z)$. We shall note this limit $K_q$ and constitutes a natural $q$-generalization of the Kolmogorov-Sinai entropy. Summarizing,
\begin{equation}
K_{q_{sen}} \equiv \lim_{t \to \infty} \lim_{W \to \infty} \lim_{N \to \infty} \frac{\langle S_{q_{sen}}\rangle (t)}{t} \;.
\end{equation}
It is easy to verify, whenever $\lambda_1 >0$, that $q_{sen}(z)=1$ and that the Pesin identity holds, i.e., $K_1 = \lambda_1$. A fascinating open question constitutes to find, whenever $\lambda_1=0$ (more specifically for $a=a_c(z)$), under what circumstances the conjecture $K_{q_{sen}} = \lambda_{q_{sen}}$ could be true, $\lambda_{q_{sen}}$ being the coefficient appearing in Eq. (15) for $q=q_{sen}$. We would then have the generalization of the Pesin identity.

Let us next address another aspect \cite{mouratirnaklilyra} concerning the edge of chaos $a_c(z)$. We spread now, at $t=0$, the $N$ points uniformly within the entire $[x_{min},x_{max}]$ interval, i.e., over {\it all} the $W$ windows, and follow, as function of time $t$, the shrinking of the number $W(t)$ of windows which contain at least one point (disappearence of the Lebesgue measure on the $x$-axis); $W(0)=W$. It can be verified that, for the sequence $ \lim_{W \to \infty} \lim_{N \to \infty}$,  we asymptotically have $W(t) \propto 1/t^{\frac{1}{q_{rel}(z)-1}}$, where $q_{eq}(z)>1$ (the subindex {\it rel} stands for {\it relaxation}). The entropic index $q_{rel}$ monotonically increases when $z$ increases from 1 to infinity;
also, within some range it is verified  \cite{mouratirnaklilyra} that $1/[q_{rel}(z)-1] \propto [1-d_f(z)]^2 $. We shall now advance a recently established \cite{borgesananostsallis} relation between $q_{rel}$ and $q_{sen}$. 

Let us go back to the procedure when, at $t=0$, only one among the $W$ windows is populated. That single window is chosen to be that which makes $S_{q_{sen}}(t)$ to achieve the highest value as $t$ increases. To be more precise $S_{q_{sen}}(0)=0$ and $0<S_{q_{sen}}(\infty) < \ln_{q_{sen}} W$. While $t$ increases, there are many windows for which $S_{q_{sen}}(t)$ overshoots above $S_{q_{sen}}(\infty)$. We choose the window for which the overshooting is the most pronounced. After $S_{q_{sen}}(t)$ achieves this peak, it relaxes slowly towards $S_{q_{sen}}(\infty)$. It does so as $1/t^{\frac{1}{q_{rel}(z,W)-1}}$, where $q_{rel}(z,W)$ approaches its limiting value $q_{rel}(z,\infty)$ while $W$ diverges. The remarkable fact is that $q_{rel}(z,\infty) = q_{rel}(z)$ ! More than this, the approach is asymptotically as follows:
\begin{equation}
q_{rel}(z) - q_{rel}(z,W) \propto \frac{1}{W^{q_{sen}(z)}}\;.
\end{equation}
This relation is a remarkable connection between the mixing properties, the equilibration (or relaxing) ones, and the degree of graining (from coarse to fine graining while $W$ increases). We may also say that in some sense Eq. (23) provides a connection between the Boltzmannian and the Gibbsian approaches to statistical mechanics. Indeed, the concept of $q_{sen}$ is kind of natural within a typical Boltzmann scenario where individual trajectories in phase space are the ``protagonists of the game", whereas $q_{rel}$ is kind of natural within a typical Gibbs scenario where the entire phase space is to be in principle occupied.  Before taking into Eq. (23) the $q_{sen}=q_{rel}=1$ particular case (i.e., the BG-like case), some adaptation is obviously needed; as written in Eq. (23), it is valid only for $q_{sen}<1$ and $q_{rel}>1$. \\

\noindent
{\it Two-dimensional maps:}

We consider here two-dimensional conservative maps of the type
\begin{eqnarray}
x_{t+1} = f_x(x_t,y_t;a,z)       \nonumber  \\
y_{t+1} = f_y(x_t,y_t;a,z)  
\end{eqnarray}
where  $x_t \in [x_{min},x_{max}]$ and $y_t \in [y_{min},y_{max}]$ with $t=0,1,2,...$; the control parameter $z$ characterizes, as for the one-dimensional maps we considered above, the universality class; the control parameter $a \ge 0$ and we assume that, while $a$ increases from zero to its maximum value, the nonnegative Lyapunov exponent $\lambda_1$ monotonically increases from zero to its maximum value. Since the map is conservative (i.e., $|\partial(x_{t+1},y_{t+1})/\partial(x_t,y_t)|=1$), the other Lyapunov exponent is $-\lambda_1$. A paradigmatic such map is the so called {\it standard map}, defined as follows
\begin{eqnarray}
y_{t+1} &=& y_t + \frac{a}{2 \pi} \sin (2\pi x_t)       \;\;\;\;\;\;\;\;\;\;\;\;\;\;\;\;\;\;\;\;\;\;\;\;\;\;\;\;\;\; (mod \; 1) \nonumber  \\
x_{t+1} &=& y_{t+1} + x_t = y_t + \frac{a}{2 \pi} \sin (2\pi x_t)   + x_t   \;\;\;(mod \;1)
\end{eqnarray}
as well as its $z$-generalization \cite{fulvioprivate}, defined as follows
\begin{eqnarray}
y_{t+1} &=& y_t + \frac{a}{2 \pi} \sin (2\pi x_t)   |\sin (2\pi x_t)|^{z-1}    \;\;\;\;\;\;\;\;\;\;\;\;\;\;\;\;\;\;\;\;\;\;\;\;\;\;\;\;\;\; (mod \; 1) \nonumber  \\
x_{t+1} &=& y_{t+1} + x_t = y_t + \frac{a}{2 \pi} \sin (2\pi x_t)   |\sin (2\pi x_t)|^{z-1} + x_t   \;\;\;(mod \;1) \;,
\end{eqnarray}
where $z \in {\cal R}$.

Some (not clearly characterized yet) classes of such maps exhibit for the entropy production $dS_q(t)/dt$ a behavior which closely follows the crossover behavior associated with Eq. (17). Let us be more precise. We first partition the accessible $(x,y)$ phase in $W$ nonoverlaping little areas (for example $W$ little squares), and put a large number $N$ of initial conditions inside one of those areas. As before, we follow along time the set of probabilities $\{p_i\}$, with which we calculate $S_q(t)$ for an arbitrarily chosen value of $q$. We then average over the entire accessible phase space and obtain $ \langle S_q \rangle (t)$. Finally we numerically approach the quantity ${\bf S}_q(t) \equiv \lim_{W \to \infty} \lim_{N \to \infty}  \langle S_q \rangle (t) $. 

For large values of $a$, we verify \cite{baldovintsallisschulze} that ${\bf S}_1(t) $ asymptotically increases linearly with $t$, as expected from the fact that $\lambda_1(a) >0$, in agreement with Pesin identity. However, an interesting phenomenon occurs for increasingly small $a$, hence increasingly small $\lambda_1$. For small $t$ (say $0 <t <<t_1(a,z)$), ${\bf S}_q(t)$ is linear with $t$ for $q=0$, and acquires an infinite slope for any $q<0$. For intermediate $t$ (say $t_1(a,z) <<t<<t_2(a,z)$), ${\bf S}_q(t)$ is linear with $t$ for $q=q_{sen}(z)<1$, acquires an infinite slope for $q<q_{sen}(z)$ and acquires a vanishing slope for  $q>q_{sen}(z)$. For large $t$ (say $t>> t_2(a,z)$), ${\bf S}_q(t)$ is linear with $t$ for $q=1$, acquires an infinite slope for $q<1$ and acquires a vanishing slope for  $q>1$. The characteristic times $t_1(a,z)$ and $t_2(a,z)$ respectively correspond to the $[q=0] \to [q=q_{sen}(z)]$ and $  [q=q_{sen}(z)] \to [q=1]$ crossovers. The remarkable feature is that, in the limit $a \to 0$, $t_1(a,z)$ remains finite whereas $t_2(a,z)$ diverges. In other words, for asymptotically small values of $a$, the time evolution of ${\bf S}_q(t)$ is, excepting for an initial transient, basically characterized by $q_{sen}(z) <1$. This fact opens the possibility for something similar to occur for Hamiltonian classical systems for which the Lyapunov spectrum tends to zero. This is precisely what occurs when the size of the system increase in the presence of long-range interactions, as we shall see in the next Subsection. Before closing this subsection, let us mention that studies focusing on $q_{rel}$ for conservative maps are in progress.

\subsection{Long-range many-body classical hamiltonian systems}

From the thermodynamical viewpoint it is interesting to classify the two-body interactions (and analogously, of course, the many-body interactions). According to their behavior near the origin, i.e., for $r \to 0$, potentials could be classified as {\it collapsing} and {\it noncollapsing}. {\it Collapsing} are those which exhibit a minimum at $r=0$. This minimum can be infinitely deep, i.e., the potential can be singular at $r=0$; such is the case of attractive potentials which asymptotically behave as $-1/r^\nu$ with $\nu>0$ (e.g., Newtonian gravitation, hence $\nu=1$). Alternatively, the potential at this $r=0$ minimum can be finite, as it is the case of those which behave as $-a + b r^{-\nu}$ with $a>0$, $b>0$ and $\nu<0$. Collapsing potentials, especially those of the singular type, are known to exhibit a variety of thermodynamical anomalies. {\it Noncollapsing} potentials are those which exhibit a minimum either at a finite distance (e.g., the Lennard-Jones one, or the hard spheres model or any other model having a cutoff at a finite distance $r_0$) or at infinity (e.g., Coulombian repulsion). Potentials can be also classified according to their behavior at $r \to \infty$. We can divide them into {\it short}- and {\it long}-range interactions. Short-range interactions are those whose associated force quickly decreases with distance, for example potentials which exponentially decrease with distance, or classical potentials of the type $-1/r^\alpha$ with $\alpha>d$, $d$ being the space dimension where the system is defined. For classical systems, thermodynamically speaking, {\it short}-range interactions correspond to the potentials which are integrable at infinity \cite{fisherlebowitz}, and {\it long}-range interactions correspond to those which are {\it not} integrable in that limit, such as those increasing like $1/r^{\alpha}$ with $\alpha<0$ (which belong to the {\it confining} class of potentials, i.e., those which make escape impossible), or those like $-1/r^\alpha$ with $0 \le \alpha/d \le1$ (which belong to the {\it nonconfining} class of potentials, i.e., those which make escape possible). Long-range interactions, especially those of the nonconfining type, are also known to induce a variety of thermodynamical anomalies. From the present standpoint a particularly complex potential is Newtonian gravitation (corresponding to $\alpha=1$ and $d=3$). Indeed, it is both singular at the origin, and long-ranged since $0<\alpha/d=1/3<1$.

In this Section we address an important case, namely that of nonsingular attractive long-range two-body interactions in a $d$-dimensional $N$-body classical hamiltonian system with $N>>1$. Such systems are being actively addressed in the literature by many authors (see \cite{HMF,anteneodotsallis} and references therein). 
As an illustration of the thermodynamical anomalies that long-range interactions produce, we shall focus on the $d$-dimensional simple hypercubic lattice with periodic boundary conditions, each site of which is occupied by a classical planar rotator. All rotators are coupled two by two as indicated in the following Hamiltonian:
\begin{equation}
{\cal H}= \sum_{i=1}^N \frac{L_i^2}{2} + \sum_{ i \ne j} \frac{1- cos(\theta_i - \theta_j)}{r_{ij}^{\;\alpha}}\;\;\;(\theta_i \in [0,2\pi];\alpha \ge0).
\end{equation}
The distance (in crystal units) between any two sites is the shortest one taking into account the periodicity of the lattice. For $d=1$, it is $r_{ij}=1,2,3,...$; for $d=2$, it is $r_{ij}=1, \sqrt 2,2,...$; for $d=3$, it is $r_{ij}=1,\sqrt 2, \sqrt 3,2,...$; and so on for higher dimensions. We have written the potential term in such a way that it vanishes in all cases for the fundamental state, i.e., $\theta_i=\theta_0 \;(\forall i)$, where, without loss of generality, we shall consider $\theta_0=0$ for simplicity. Also without loss of generality we have considered unit moment of inertia and unit first-neighbor coupling constant. It is clear that, excepting for the inertial term, the present model is nothing but the classical $XY$ ferromagnet. The cases $\alpha=0$ and $\alpha \to \infty$ respectively correspond to the so called HMF model \cite{HMF} (all two-body couplings have the same strength), and to the first-neighbor model. For BG  statistical mechanics to be applicable without further considerations, it is necessary that the potential be integrable, i.e., $\int_1^\infty dr\; r^{d-1} r^{-\alpha} <\infty$. This implies $\alpha >d$. In this case, the energy of the system is extensive, i.e., the energy per particle is finite in the thermodynamic limit $N \to \infty$. But the situation becomes more delicate for $0 \le \alpha/d \le 1$, since then that integral diverges. However, strictly speaking, the system being finite, the integral that is to be analyzed is not the one already mentioned but the following one instead:
\begin{equation}
\int_1^{N^{1/d}} dr \;r^{d-1} r^{-\alpha},
\end{equation}
which, in the $N \to \infty$ limit, converges for $\alpha/d>1$ and diverges otherwise. It is in fact convenient to introduce the quantity
\begin{equation}
{\tilde N} \equiv 1+d\int_1^{N^{1/d}} dr \;r^{d-1} r^{-\alpha} = \frac{N^{1-\alpha/d}-\alpha/d}{1-\alpha/d}\;.
\end{equation}
${\tilde N}$ equals $N$ for $\alpha=0$, and, for $N \to \infty$, diverges like $N^{1-\alpha/d}$ for $0<\alpha/d$, diverges like $\ln N$ for $\alpha/d=1$, and is finite for $\alpha/d>1$, being unity in the limit $\alpha/d \to \infty$ . In general, the energy per particle scales with ${\tilde N}$; in other words, the energy is nonextensive for $0 \le \alpha/d \le 1$. To make the problem artificially extensive even for $\alpha/d \le 1$, the Hamiltonian can be written as follows:
\begin{equation}
{\cal H^{\prime}}= \sum_{i=1}^N \frac{L_i^2}{2} + \frac{1}{{\tilde N}}\sum_{ i \ne j} \frac{1- cos(\theta_i - \theta_j)}{r_{ij}^{\;\alpha}}\;.
\end{equation}
The rescaling of the potential of this model is more properly taken into account by $\sum_{ i \ne j} r_{ij}^{-\alpha}$ \cite{panchocelia} rather than by ${\tilde N}$, but since their $N \to \infty$ asymptotic behaviors coincide, we can as well use ${\tilde N}$ as introduced here. The original (Eq. (27)) and rescaled (Eq. (30)) versions of this model are completely equivalent (see \cite{anteneodotsallis}) and lead to results that can be easily transformed from one to the other version. To make easier the comparison of results existing in the literature, we shall from now on refer to the rescaled version (30). 

The $\alpha=0$ model (HMF) clearly is $d$-independent and is paradigmatic of what happens for any $\alpha$ such that $0 \le \alpha/d <1$. When isolated (microcanonical ensemble) the $\alpha/d=0$ model exhibits a second-order phase transition at $u \equiv U/N = 0.75$, where $U$ is its total energy and $N \to \infty$. This critical value $u_c$ smoothly increases with $\alpha/d$ approaching unity. Dynamical and thermodynamical anomalies exist in both ordered and disordered phases, respectively for $u<u_c$ and $u>u_c$. Let us discuss some anomalies for $u>u_c$, then some for $u<u_c$, and finally show that these anomalies on both sides of $u_c$ are in fact connected. 

The Lyapunov spectrum is made by couples of real quantities that are equal in absolute value and opposite in sign, whose sum vanishes in accordance with the Liouville theorem. The sum of the positive values equals the Kolmogorov-Sinai entropy, in accordance with the Pesin theorem. If the maximal Lyapunov exponent vanishes, the entire spectrum vanishes, and no exponentially quick sensitivity to the initial conditions is possible. 

We address first the case $u>u_c$. For the Hamiltonian of rotors we are interested in (Eq. (30)), the maximal Lyapunov exponent ${\tilde \lambda}^{max}$ scales, for large $N$, like
\begin{equation}
{\tilde \lambda}^{max} \sim \frac{l(u,\alpha,d)}{N^{\kappa(\alpha/d)}} \;,
\end{equation}
where $l(u, \alpha,d)$ is some smooth function of its variables, and $\kappa(\alpha/d)$ decreases from 1/3 to zero when $\alpha/d$ increases from zero to unity; $\kappa$ remains zero for all values of $\alpha/d$ above unity, consistently with the fact that ${\tilde \lambda}^{max}$ is positive in that region. In other words, above the critical energy, the sensitivity to the initial conditions is exponential for $\alpha/d>1$, and subexponential (possibly power like) for $0 \le \alpha/d \le 1$. 

We address now the case $0<u<u_c$, and focus especially on the region slightly below the critical value $u_c$ (e.g., $u=0.69$ for the $\alpha=0$ model). For $0 \le \alpha/d \le 1$, at least two (and possibly only two with nonzero measure) important basins exist in the space of the initial conditions: one of them contains the Maxwellian distribution of velocities, the other one contains the water-bag (as well as the double water-bag) distibution of velocities. When the initial conditions belong to the Maxwellian basin, the system relaxes quickly onto the BG equilibrium distribution (strictly speaking, we do not have this numerical evidence but a weaker one, namely that the marginal probability of one-rotator velocities tends to the Maxwellian one when $N \to \infty$). When the initial conditions belong to the other basin, it first relaxes quickly to an anomalous, metastable (quasi-stationary) state, and only later, at a crossover time $\tau$, starts slowly approaching the BG equilibrium. The crossover time diverges with $N$ for $\alpha=0$ \cite{latorarapisardatsallis}. It has been conjectured \cite{tsallisconjecture} that it might in general diverge like $\tau \sim {\tilde N}$. It has been recently established \cite{giansantietal} that, for $d=1$ and fixed $N$, $\tau$ exponentially vanishes with $\alpha$ approaching unity. All these features are consistent with the conjecture $\tau \sim {\tilde N}$, which might well be true. During the metastable state, the one-particle distribution of velocities is clearly non Gaussian, and in fact it seems to approach the distribution of velocities typical of nonextensive statistical mechanics for $q>1$. This anomalous behavior reflects on the sensitivity to the initial conditions. The maximal Lyapunov exponent ${\tilde \lambda}^{max}$ remains during long time, in fact until $t \sim \tau$, at a low value and then starts approaching a finite value. This low value scales like $1/N^{\kappa^\prime(\alpha/d)}$. The remarkable feature which has been observed \cite{andreacagliari,bene} for the $d=1$ model is that $\kappa^\prime = \kappa/3$ for all values of $\alpha$. The anomalies above and below the critical point become thus intimately related. 

The whole scenario is expected to hold for large classes of models, including the classical $n$-vector ferromagnetic-like two-body coupled inertial rotors ($n=2$ being the present one, $n=3$ the Heisenberg one, $n \to \infty$ the spherical one, etc). For all of them, in the isolated situation, we expect (i) at the disordered phase, that the maximal Lyapunov decreases with $N$ with the exponent $\kappa(\alpha/d,n)$ (it is yet unclear whether this exponent depends on $n$ or not); (ii) at the ordered phase, and starting from initial conditions within a finite basin including the water-bag, that a metastable state exists with non BG (possibly $q$-type) distribution associated with a maximal Lyapunov exponent which decreases with $N$ with the exponent $\kappa(\alpha/d)/3$. In these circumstances, for $\alpha/d \le 1$ (nonextensive systems), the $\lim_{N \to \infty} \lim_{t \to \infty}$ ordering is expected to yield the usual BG equilibrium, whereas the $ \lim_{t \to \infty} \lim_{N \to \infty}$ ordering yields a non BG (meta)equilibrium, possibly of the type predicted by nonextensive statistical mechanics. This interesting phenomenon disappears for $\alpha/d > 1$ (extensive systems); indeed, both orderings lead then to the same equilibrium, namely the BG one, as known since long.

\section{Conclusions}

We have presented some of the main peculiarities associated with nonextensive systems. Most of the paradigmatic behaviors are expected to become (or have been shown to become) power laws instead of the usual exponentials:

(i) the sensitivity to the initial conditions is given by $\xi = e_q^{\lambda_q\; t}$ (typically $q \le 1$); 

(ii) the finite entropy production (Kolmogorov-Sinai entropy like) occurs only for $S_q$ (with $q\le 1$, the same as above);

(iii) the relaxation towards quasi-stationary or (metaequilibrium) states, or perhaps from these to the terminal equilibrium states, may occur through $e_q^{-t/\tau_q}$ (typically $q \ge 1$);

(iv) the stationary, (meta)equilibrium distribution for thermodynamically large Hamiltonian systems may be given by $p_i \propto e_q^{-\beta^\prime \epsilon_i}$ (typically $q \ge 1$, possibly the same as just above).

The two-dimensional conservative maps exhibit, in the vicinity of integrability and at intermediate times, features very similar to those observed in one-dimensional dissipative maps at the edge of chaos. The intermediate stage has a duration which diverges when the control parameters approach values where the system is close to integrability. Isolated classical Hamiltonian systems behave similarly to low-dimensional conservative maps, $1/N$ playing a role analogous to the distance of the control parameters to their values where integrability starts.

The scenario which emerges is that sensitivity and entropy production properties are related to one and the same value of $q_{sen} \le 1$ (also related to $\alpha_{min}$ and $\alpha_{max}$ of some multifractal function), whereas the relaxation and (meta)equilibrium properties are related to (possibly one and the same) value of $q_{rel} \ge 1$ (also related to the Hausdorff dimension of the same multifractal function). These two sets of properties are quite distinct and generically correspond to distinct values of $q$ (namely, $q_{sen}$ and $q_{rel}$). It happens that for the usual, extensive, BG systems they coincide providing $q_{sen}=q_{rel}=1$, which might sometimes be at the basis of some confusion. In all cases, once the microscopic dynamics of the systems is known, it is in principle possible to determine {\it a priori} both $q_{sen}$ and $q_{rel}$ (as well as the connection among them and with the chosen graining, as illustrated in Eq. (23)). We have here shown how this is done for simple systems. 
This type of calculation of $q$ from first principles has also been illustrated for a variety of other systems \cite{multiplicative}.
The quest for such calculations for more complex systems is in progress.

\section*{Acknowledgements}

I thank I. Bediaga, E.M.F. Curado, J. Miranda, C. Beck, T. and N. Arimitsu,  A. Upadhyaya, J.-P. Rieu, J.A. Glazier and Y. Sawada and M.A. Montemurro
for kindly providing and allowing me to use the figures that are shown here. I also thank CNPq, PRONEX and FAPERJ (Brazilian agencies) and FCT (Portugal) for partial
financial support. Finally, I am grateful to B.J.C. Cabral, who provided warm hospitality at the Centro de Fisica da Materia Condensada of Universidade de Lisboa, where this work was partially done.

\begin{figure}
\caption{The distributions of differences of velocity in two different experiments of fluid turbulence. The solid lines correspond to Beck' s theory \protect\cite{beck2}.}  
\label{fig1}
\end{figure}

\begin{figure}
\caption{The distributions of differences of velocity in a numerical experiments of fluid turbulence, for typical values of $r/\eta$. The solid lines correspond to the Arimitsu's theory (see \protect\cite{arimitsu2} for details). The entire set of theoretical curves has been obtained with a single value $q_{sen} <1$ .} 
\label{fig2}
\end{figure}

\begin{figure}
\caption{Distributions of transverse momenta of hadronic jets produced in electron-positron annihilation. The solid lines correspond to the Bediaga-Curado-Miranda theory \protect\cite{epjet}. To each curve, a different value of $q$ (in the range $(1,1.2)$) is associated. The dashed line corresponds to Hagedorn's theory using BG statistics ($q=1$).} 
\label{fig3}
\end{figure}

\begin{figure}
\caption{Anomalous diffusion measurements of cells of {\it Hydra viridissima} \protect\cite{hydra}. The dot-dashed line corresponds to normal diffusion ($q=1$), whereas the solid line corresponds to anomalous superdiffusion associated with $q=1.5$ .} 
\label{fig4}
\end{figure}

\begin{figure}
\caption{Distribution of velocities of cells of {\it Hydra viridissima} \protect\cite{hydra}. The solid line corresponds to $q=1.5$, thus coinciding with the value of Fig. 4.} 
\label{fig5}
\end{figure}

\begin{figure}
\caption{Zipf plot (frequency of words with rank $r$) associated with various books as indicated on the figure ($N$ is the total number of words; $V$ is the vocabulary, i.e., the number of different words). See details in \protect\cite{montemurro}.} 
\label{fig6}
\end{figure}

\begin{figure}
\caption{Zipf plot associated with 56 books by Charles Dickens. The solid line corresponds to $q=1.9$, and the crossover to the $q=1$ regime at high rank $r$ is visible on the figure. See details in \protect\cite{montemurro}.} 
\label{fig7}
\end{figure}


\references


\bibitem{tsallis}
C. Tsallis, J. Stat. Phys. {\bf 52}, 479 (1988).

\bibitem{curadotsallis}E.M.F. Curado and C. Tsallis, J. Phys. A {\bf 24}, L69 (1991) [Corrigenda: {\bf 24}, 3187 (1991) and {\bf 25}, 1019 (1992)].

\bibitem{tsallismendesplastino}C. Tsallis, R.S. Mendes and A.R. Plastino, Physica A {\bf 261}, 534 (1998).

\bibitem{review}S.R.A. Salinas and C. Tsallis, {\it Nonextensive Statistical Mechanics and Thermodynamics}, Braz. J. Phys. {\bf 29}, Number 1 (1999);
S. Abe and Y. Okamoto, {\it Nonextensive Statistical Mechanics and Its Applications}, Series {\it Lectures Notes in Physics} (Springer, Berlin, 2001);
P. Grigolini, C. Tsallis and B.J. West, {\it Classical and Quantum Complexity and Nonextensive Thermodynamics}, Chaos, Solitons and Fractals {\bf 13}, Issue 3 (2002);
G. Kaniadakis, M. Lissia and A. Rapisarda, {\it Non Extensive Thermodynamics and Physical Applications}, Physica A {\bf 305}, 1/2 (2002); M. Gell-Mann and C. Tsallis, {\it Nonextensive Entropy - Interdisciplinary Applications} (Oxford University Press, Oxford, 2002), in preparation.

\bibitem{biblio}http://tsallis.cat.cbpf.br/biblio.htm

\bibitem{levy}
C. Tsallis, S.V.F. Levy, A.M.C. Souza  and R. Maynard, Phys. Rev. Lett.  {\bf 75}, 3589 (1995) [Erratum: {\bf 77}, 5442 (1996)];
D.H. Zanette and P.A. Alemany, {Phys. Rev. Lett.} {\bf 75}, 366 (1995); 
M.O. Caceres and C.E. Budde, Phys. Rev. Lett. {\bf 77}, 2589  (1996); D.H. Zanette and P.A. Alemany, Phys. Rev. Lett. {\bf 77}, 2590 (1996); 
M. Buiatti, P. Grigolini, and A. Montagnini, Phys. Rev. Lett. {\bf 82}, 3383 
(1999); D. Prato and C. Tsallis, Phys. Rev. E
{\bf 60}, 2398 (2000).

\bibitem{perbak}P. Bak, {\it How Nature Works} (Springer-Verlag, New York, 1996).

\bibitem{mandelbrot}B.M. Mandelbrot, {\it The Fractal Geometry of Nature} (W.H. Freeman, New York, 1982).

\bibitem{gellmann}M. Gell-Mann, {\it The Quark and the Jaguar: Adventures in the Simple and the Complex} (W.H. Freeman, New York, 1999). 

\bibitem{ramos}F.M. Ramos, R.R. Rosa and C. Rodrigues Neto, cond-mat/9907348 (1999); F.M. Ramos, C. Rodrigues Neto and R.R. Rosa, cond-mat/0010435 (2000); F.M. Ramos, R.R. Rosa, C. Rodrigues Neto, M.J.A. Bolzan, L.D.A. Sa and H.F. Campos Velho, Physica A {\bf 295}, 250 (2001); C. Rodrigues Neto, A. Zanandrea, F.M. Ramos, R.R. Rosa, M.J.A. Bolzan and L.D.A. Sa, Physica A {\bf 295}, 215 (2001); H.F. Campos Velho, R.R. Rosa, F.M. Ramos, R.A. Pielke, G.A. Degrazia, C. Rodrigues Neto and Z. Zanandrea, Physica A {\bf 295}, 219 (2001).

\bibitem{beck}C. Beck, Physica A {\bf 277}, 115 (2000); Phys. Lett. A {\bf 287}, 240 (2001); B.K. Shivamoggi and C. Beck,  J. Phys. A {\bf 34}, 4003 (2001); C. Beck, G.S. Lewis and H.L. Swinney, Phys. Rev. E {\bf 63}, 035303 (2001).

\bibitem{arimitsu}T. Arimitsu and N. Arimitsu, Phys. Rev. E {\bf 61}, 3237 (2000); J. Phys. A {\bf 33}, L235 (2000) [Corrigenda: {\bf 34}, 673 (2001)]; Physica A {\bf 295}, 177 (2001). 

\bibitem{epjet}I. Bediaga, E.M.F. Curado and J. Miranda,  Physica A {\bf 286}, 156 (2000); C. Beck, Physica A {\bf 286}, 164 (2000).

\bibitem{hydra}A. Upadhyaya, J.-P. Rieu, J.A. Glazier and Y. Sawada,  Physica A {\bf 293}, 549 (2001).

\bibitem{montemurro}M.A. Montemurro, Physica A {\bf 300}, 567 (2001).

\bibitem{boghosian}B.M. Boghosian, Phys. Rev. E {\bf 53}, 4754 (1996).

\bibitem{beck2}C. Beck, Phys. Rev. Lett. {\bf 87}, 180601 (2001).

\bibitem{arimitsu2}T. Arimitsu and N. Arimitsu, Physica A {\bf 305}, 218 (2002).               

\bibitem{bukman}C. Tsallis and D.J. Bukman, Phys. Rev. E {\bf 54}, R2197 (1996).

\bibitem{denisov}S. Denisov, Phys. Lett. A {\bf 235}, 447 (1997).

\bibitem{plastinospolytrope}A.R. Plastino and A. Plastino, Phys. Lett. A {\bf 174}, 384 (1993).

\bibitem{maps1}C. Tsallis, A.R. Plastino and W.-M. Zheng, Chaos, Solitons and Fractals {\bf 8}, 885 (1997); U.M.S. Costa, M.L. Lyra, A.R. Plastino and C. Tsallis, Phys. Rev. E {\bf 56}, 245 (1997); M.L. Lyra, Ann. Rev. Comp. Phys. , ed. D. Stauffer (World Scientific, Singapore, 1998), page 31; U. Tirnakli, C. Tsallis and M.L. Lyra, Eur. Phys. J. B  {\bf 11}, 309 (1999); U. Tirnakli, Phys. Rev. E {\bf 62}, 7857 (2000); C.R. da Silva, H.R. da Cruz and M.L. Lyra, in {\it Nonextensive Statistical Mechanics and Thermodynamics}, eds. S.R.A. Salinas and C. Tsallis, Braz. J. Phys. {\bf 29}, 144 (1999); U. Tirnakli, G.F.J. Ananos and C. Tsallis, Phys. Lett. A {\bf 289}, 51 (2001); J. Yang and P. Grigolini, Phys. Lett. A {\bf 263}, 323 (1999).

\bibitem{fulvioalberto1}F. Baldovin and A. Robledo, cond-mat/0205356.

\bibitem{fulvioalberto2}F. Baldovin and A. Robledo, cond-mat/0205371.

\bibitem{lyratsallis}M.L. Lyra and C. Tsallis, Phys. Rev. Lett. {\bf 80}, 53 (1998).

\bibitem{latorabarangerrapisardatsallis}V. Latora, M. Baranger, A. Rapisarda and C. Tsallis, Phys. Lett. A {\bf 273}, 97 (2000).

\bibitem{mouratirnaklilyra}F.A.B.F. de Moura, U. Tirnakli and M.L. Lyra, Phys. Rev. E {\bf 62}, 6361 (2000).

\bibitem{borgesananostsallis}E.P. Borges, C. Tsallis, G.F.J. Ananos and P.M.C. Oliveira, cond-mat/0203348.

\bibitem{fulvioprivate}F. Baldovin and C. Tsallis, unpublished (2001).

\bibitem{baldovintsallisschulze}F. Baldovin, C. Tsallis and B. Schulze, cond-mat/0108501 (2001).

\bibitem{HMF}M. Antoni and S. Ruffo, Phys. Rev. E {\bf 52}, 2361 (1995).

\bibitem{anteneodotsallis}C. Anteneodo and C. Tsallis, Phys. Rev. Lett. {\bf 80}, 5313 (1998).

\bibitem{fisherlebowitz}M.E. Fisher, Arch. Rat. Mech. Anal. {\bf 17}, 377 
(1964); J. Chem. Phys. {\bf 42}, 3852 
(1965); J. Math. Phys. {\bf 6}, 1643 (1965); M.E. Fisher and D. Ruelle, 
J. Math. Phys. {\bf 7}, 260 
(1966); M.E. Fisher and J.L. Lebowitz, Commun. Math. Phys. {\bf 19}, 
251 (1970).

\bibitem{panchocelia}F.A. Tamarit and C. Anteneodo, Phys. Rev. Lett. {\bf 84}, 208 (2000).

\bibitem{tsallisconjecture}C. Tsallis, communicated at the HMF Meeting (Universita di Catania, 6-8 september 2000).

\bibitem{latorarapisardatsallis}V. Latora, A. Rapisarda and C. Tsallis, Phys. Rev. E {\bf 64}, 056134 (2001).

\bibitem{giansantietal}A. Campa, A. Giansanti and D. Moroni, Physica A {\bf 305}, 137 (2002).

\bibitem{andreacagliari}V. Latora, A. Rapisarda and C. Tsallis, Physica A {\bf 305}, 129 (2002).

\bibitem{bene}B.J.C. Cabral and C. Tsallis, cond-mat/0204029.

\bibitem{multiplicative}C. Anteneodo and C. Tsallis, cond-mat/0205314.
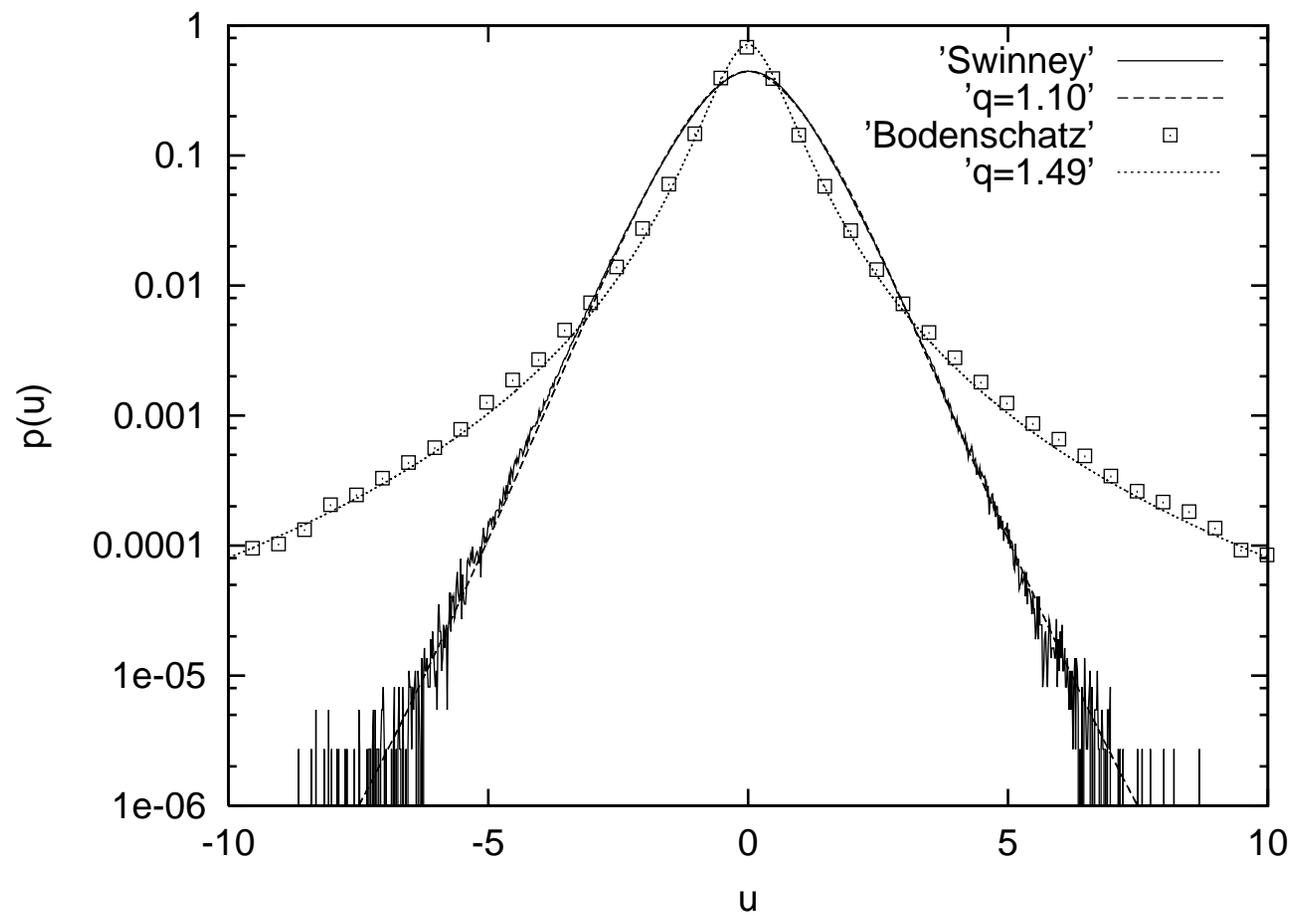
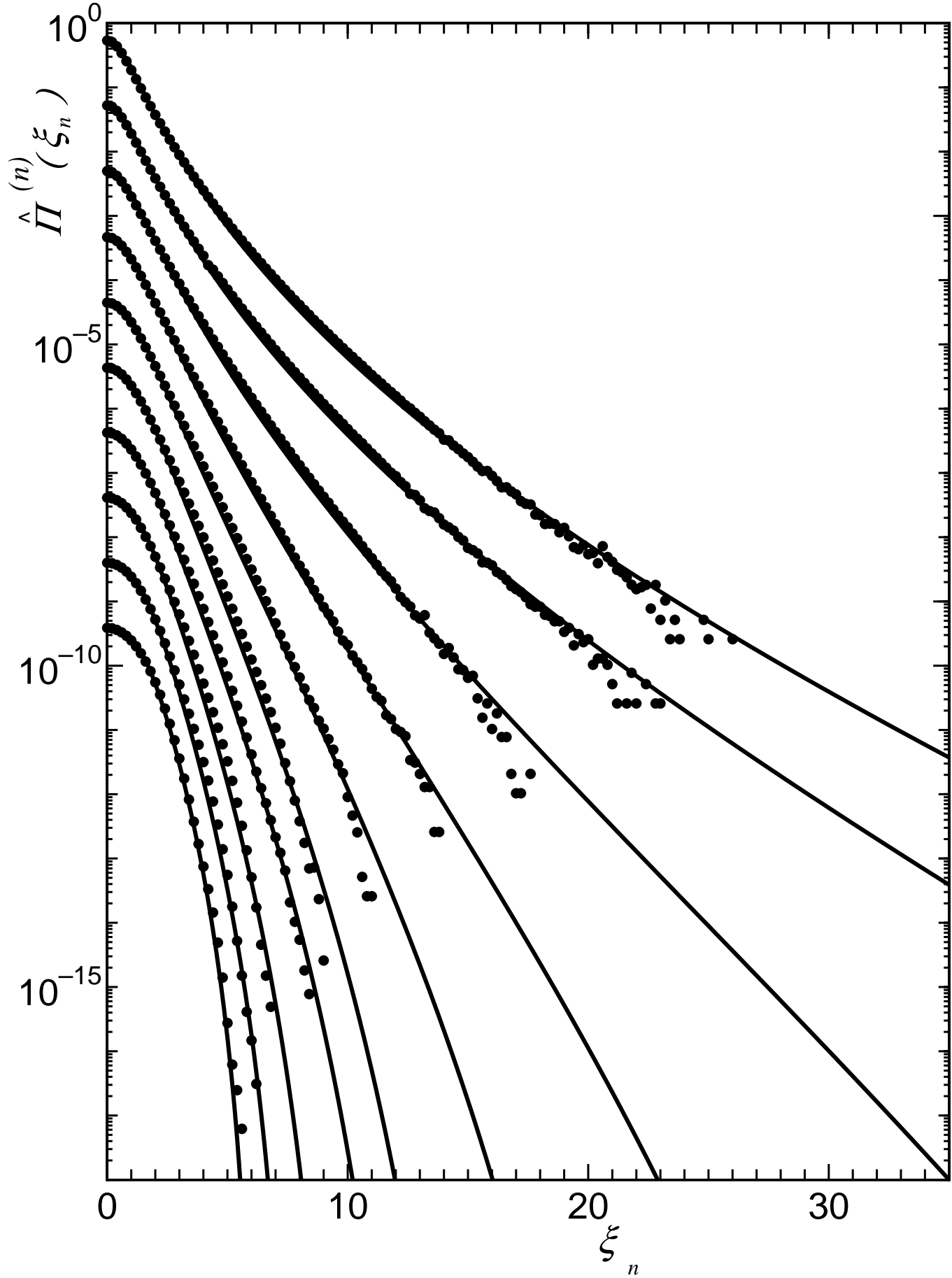
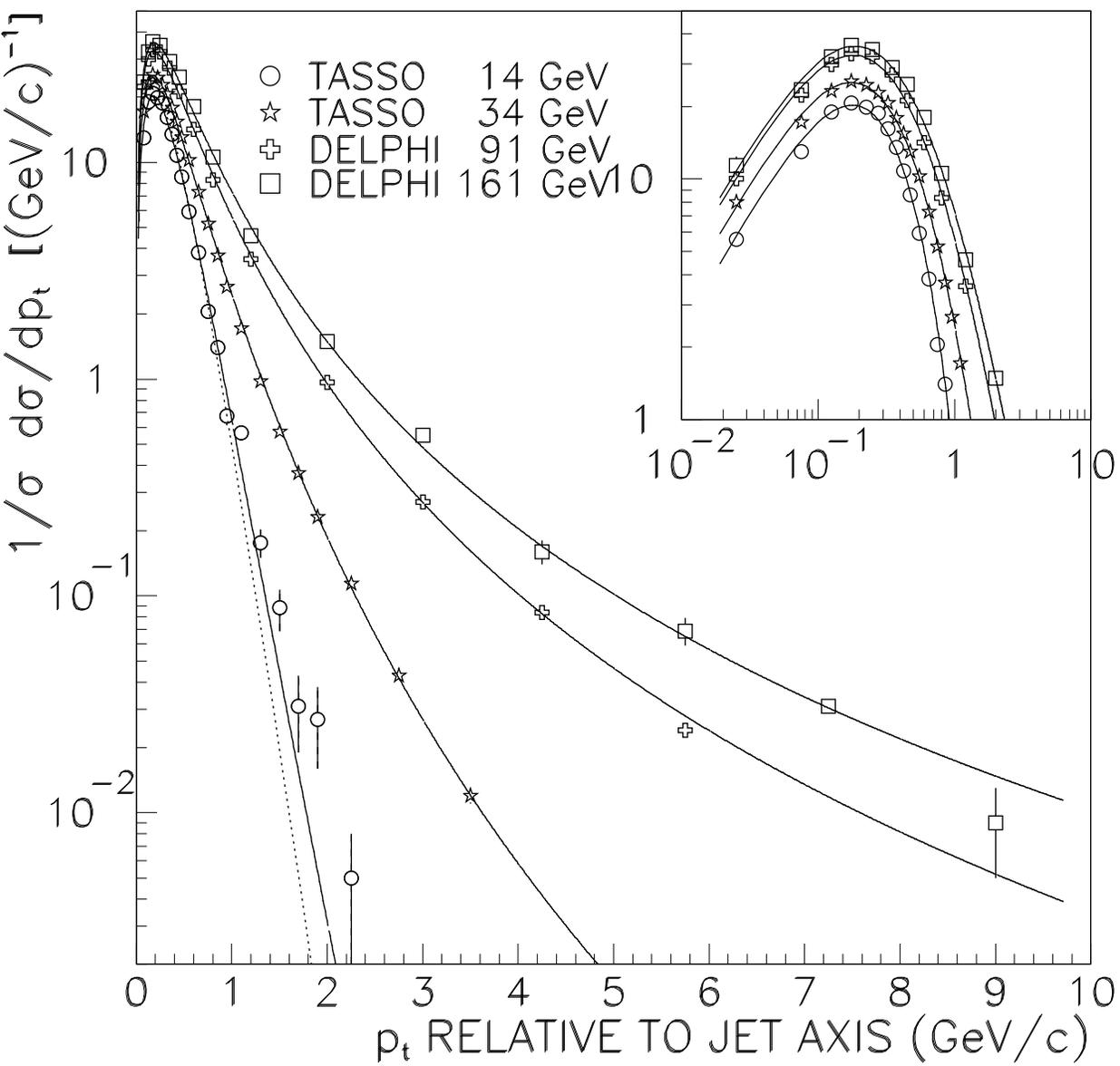
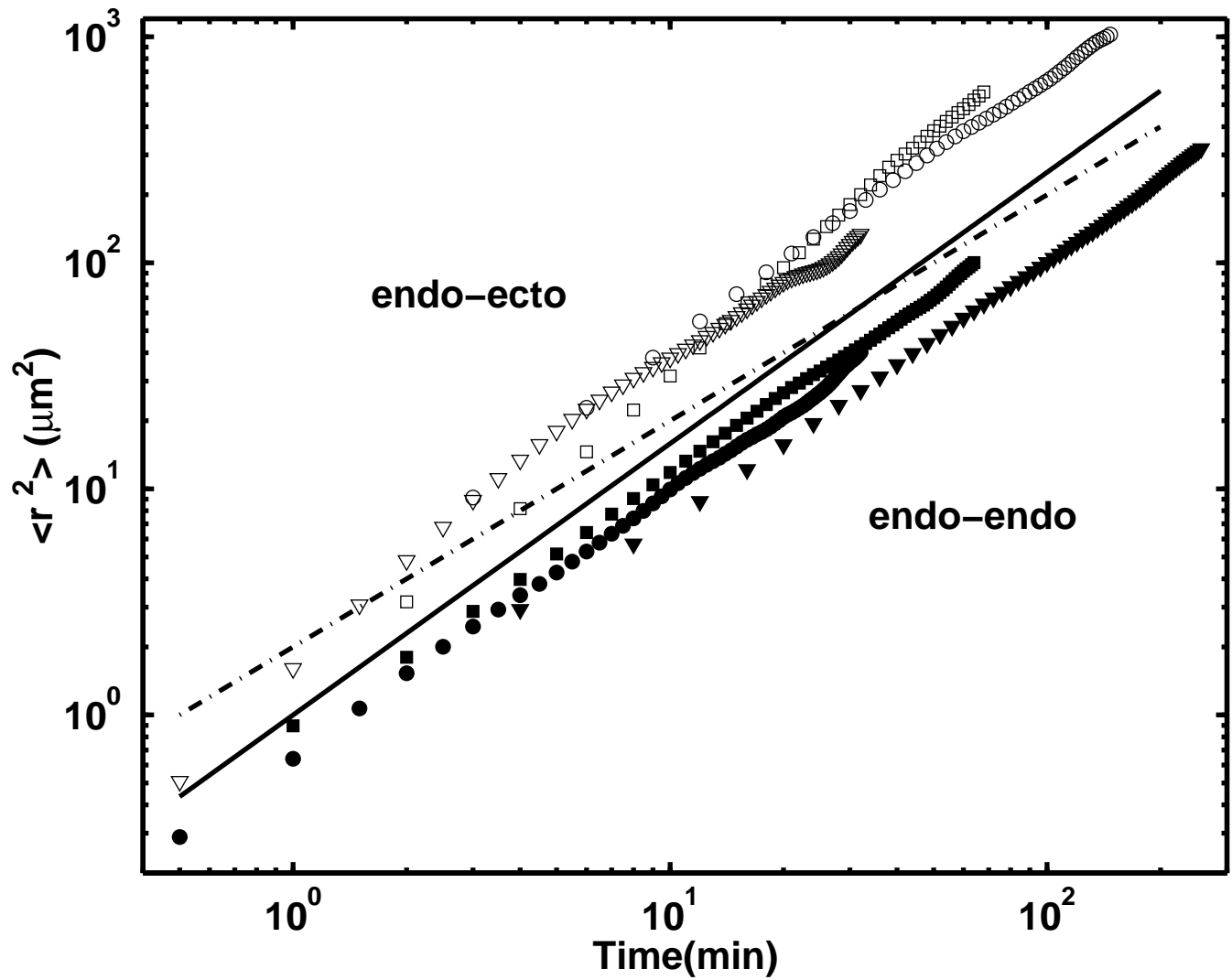
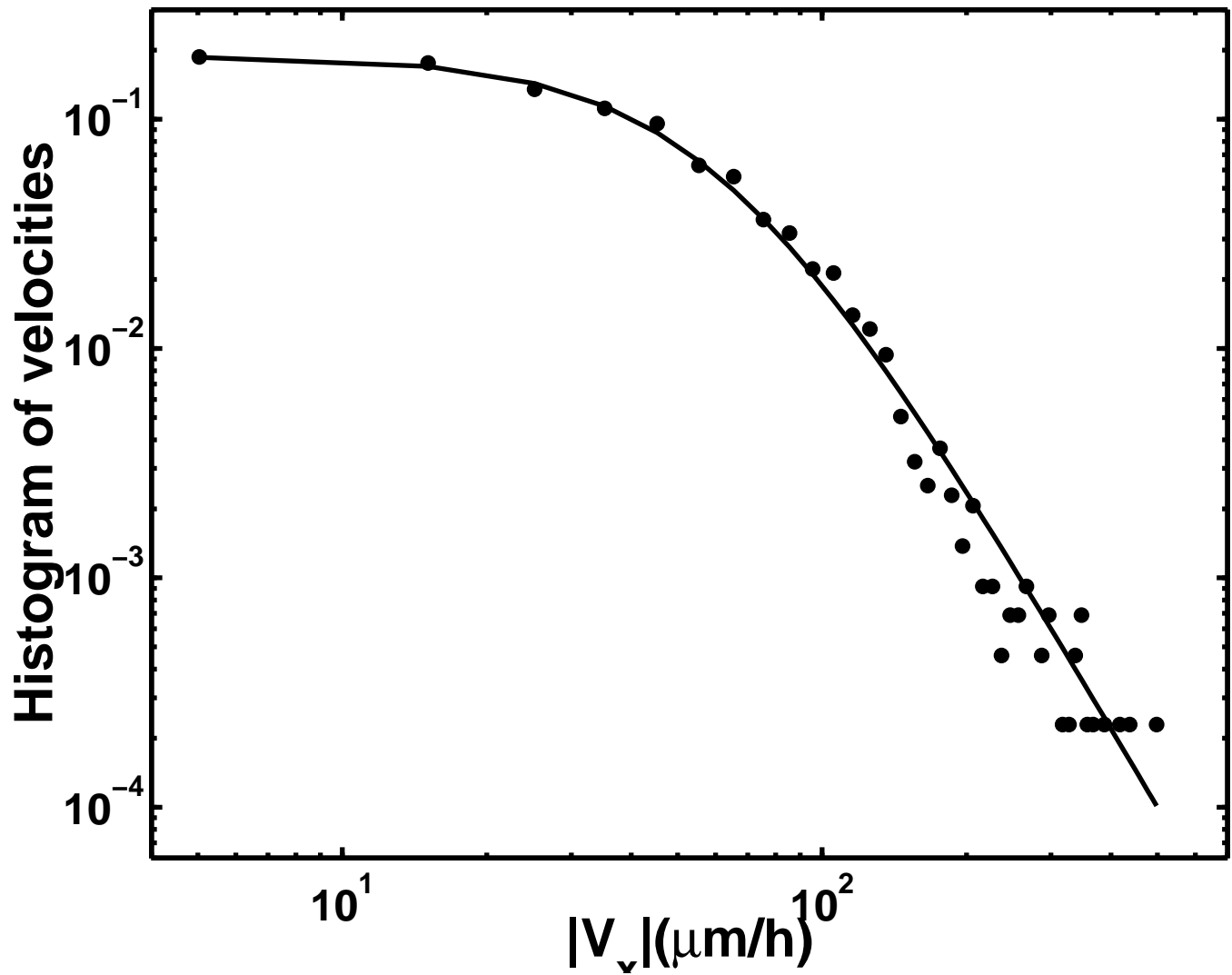
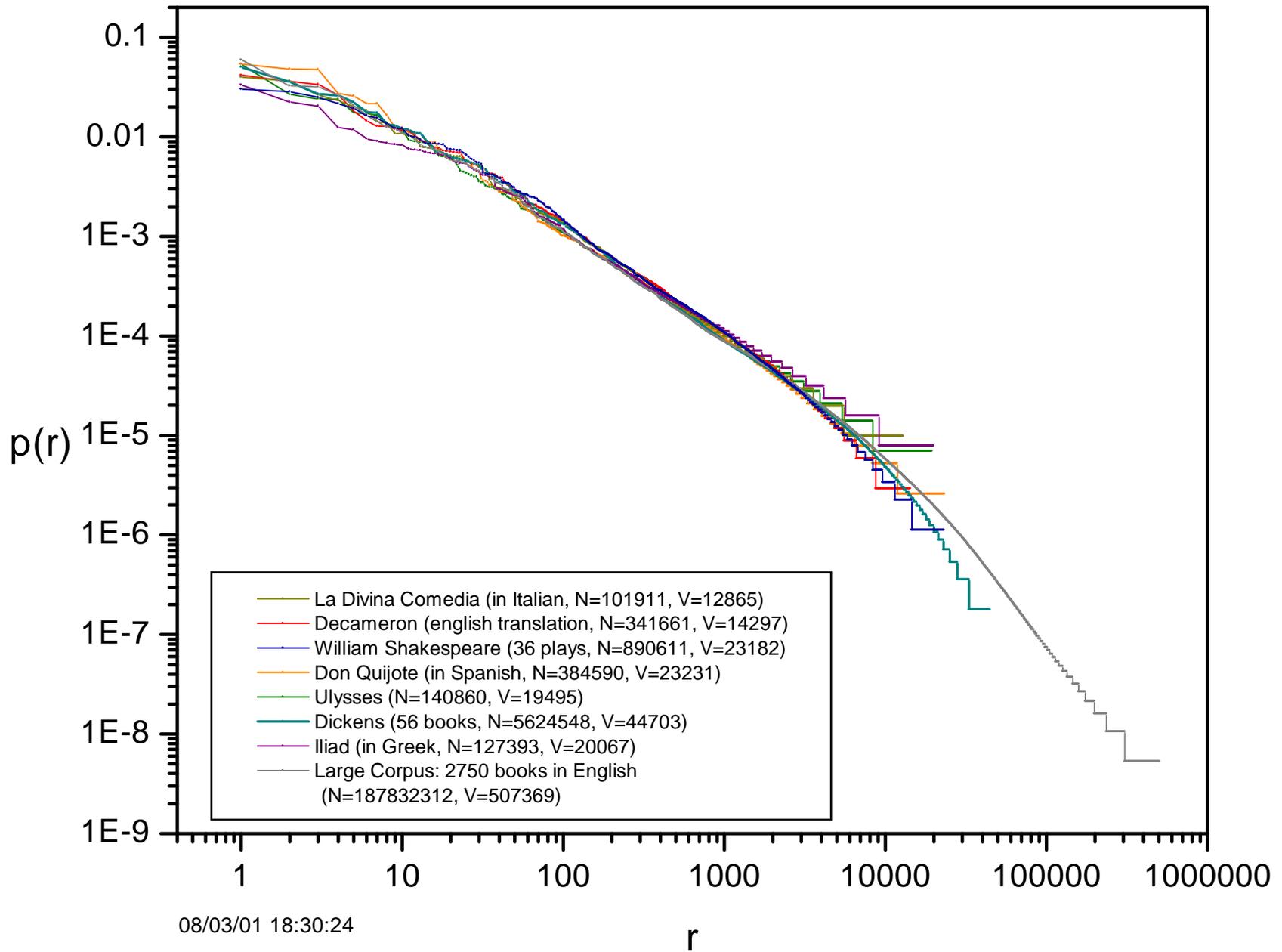
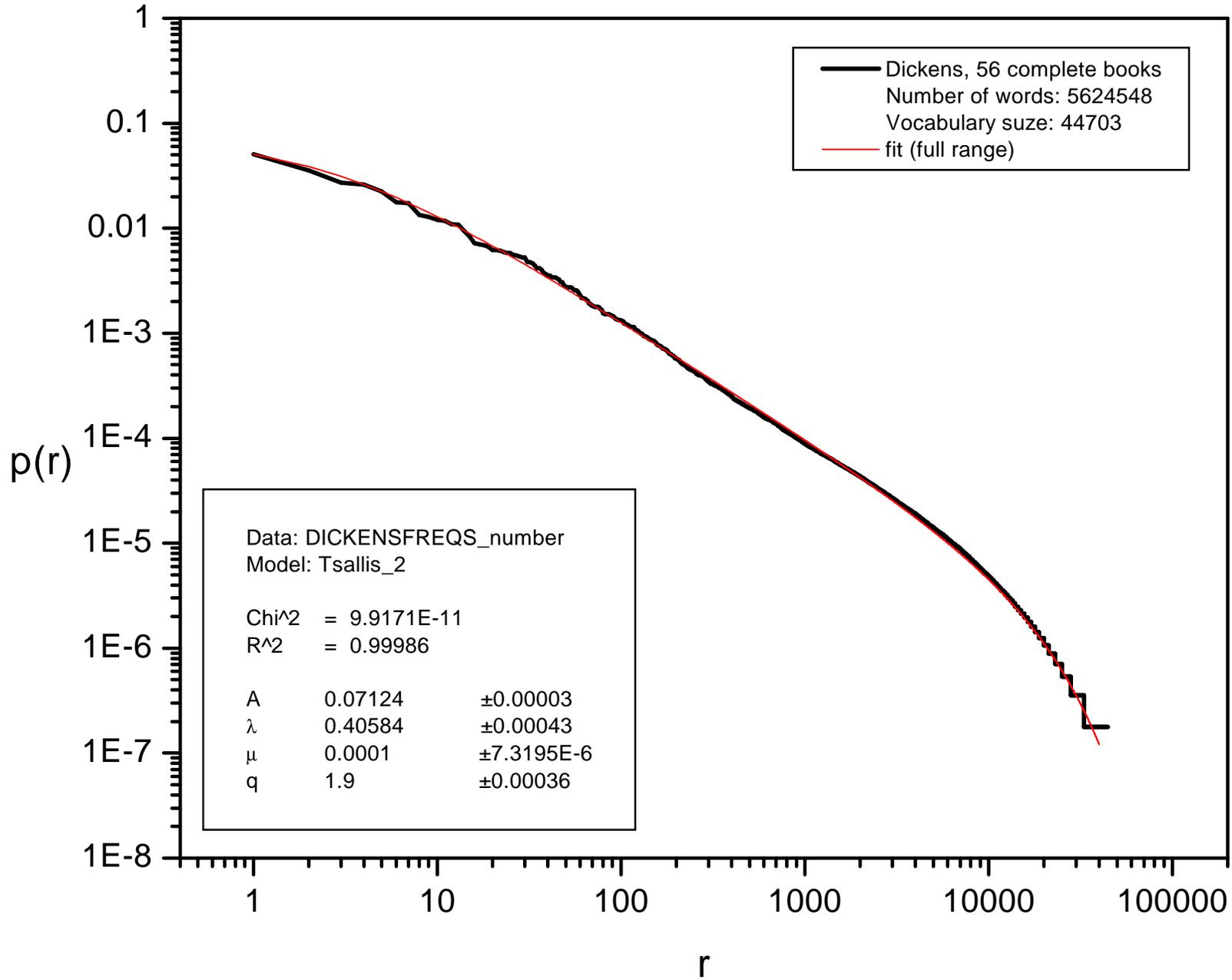
\end{document}